\documentstyle[12pt]{article}

\newcommand{\EQ}{\begin{equation}}
\newcommand{\EN}{\end{equation}}

\newcommand{\bear}{\begin{eqnarray}}
\newcommand{\ear}{\end{eqnarray}}

\begin{document}

\topmargin 0pt
\oddsidemargin 5mm
\newcommand{\NP}[1]{Nucl.\ Phys.\ {\bf #1}}
\newcommand{\PL}[1]{Phys.\ Lett.\ {\bf #1}}
\newcommand{\NC}[1]{Nuovo Cimento {\bf #1}}
\newcommand{\CMP}[1]{Comm.\ Math.\ Phys.\ {\bf #1}}
\newcommand{\PR}[1]{Phys.\ Rev.\ {\bf #1}}
\newcommand{\PRL}[1]{Phys.\ Rev.\ Lett.\ {\bf #1}}
\newcommand{\MPL}[1]{Mod.\ Phys.\ Lett.\ {\bf #1}}
\newcommand{\JETP}[1]{Sov.\ Phys.\ JETP {\bf #1}}
\newcommand{\TMP}[1]{Teor.\ Mat.\ Fiz.\ {\bf #1}}
     
\renewcommand{\thefootnote}{\fnsymbol{footnote}}
     
\newpage
\setcounter{page}{0}
\begin{titlepage}     
\begin{flushright}
\end{flushright}
\vspace{0.5cm}
\begin{center}
\large{ Applications of Temperley-Lieb algebras to Lorentz lattice gases } \\
\vspace{1cm}
\vspace{1cm}
 {\large M. J.  Martins$^{1}$ ,  B. Nienhuis$^{2}$ } \\
\vspace{1cm}
\centerline{\em ${}^{1}$ Departamento de F\'isica, 
Universidade Federal de S\~ao Carlos}
\centerline{\em Caixa Postal 676, 13565-905, S\~ao Carlos, Brazil}
\centerline{ \em and }
\centerline{\em ${}^{2}$ Instituut voor Theoretische Fysica, 
Universiteit van Amsterdam }
\centerline{\em  Valcknierstraat 65, 1018 XE Amsterdam, The Netherlands}
\vspace{1.2cm}   
\end{center} 
\begin{abstract}
Motived by the study of motion in a random environment we introduce 
and investigate a variant of the Temperley-Lieb algebra. This algebra
is very rich,
providing  us three classes of solutions of the Yang-Baxter equation. This
allows us to 
establish a theoretical
framework to study the diffusive behaviour of a Lorentz Lattice gas. 
Exact results 
for the geometrical scaling behaviour
of closed paths are also presented.
\end{abstract}
\vspace{.2cm}
\centerline{PACS numbers: 05.50.+q, 0.5.20.-y, 04.20.Jb, 11.25.Hf }
\vspace{.2cm}
\centerline{July 1998}
\end{titlepage}

\renewcommand{\thefootnote}{\arabic{footnote}}

The Temperley-Lieb(TL) algebra arosed in the context of statistical
mechanics and it was used to map the $q$-state Potts models into
the 6-vertex model \cite{TL}. The relevance of this algebra to the
theory of two-dimensional solvable models was further elaborated
by Baxter \cite{BA} and the first detailed study of its mathematical
structure was given by Jones \cite{JO}. In recent years, the 
interest in TL algebras is widespread in many branches of physics
and mathematics, including quantum spin chains \cite{BAT,LE},
conformal field theory \cite{PAS} and knot theory \cite{KAU,WA} just
to name a few.

In this letter, motived by physical considerations, we introduce and 
investigate a variant of the TL algebra. Our physical motivation is to
describe the diffusion behaviour of a Lorentz lattice gas, consisting of a
dilute gas of particles moving in a random array of scatterers \cite{CO}
. In the
model we shall consider here, a particle moves along the bonds of the
square lattice and its trajectory is changed when it hits scatterers
which are either (right and left) mirrors or rotators randomly placed
at the sites of the lattice \cite{CO,CO1}. This 
is perhaps the simplest nontrivial
``scattering rules'' one can think of  and they
are illustrated in Fig.1 . To describe the statistical behaviour of these
paths in a more general way, we assign a fugacity $q$ to every 
closed trajectory. This then defines a loop model and its 
partition function is given by
\EQ
Z= \sum_{ scatter~ configurations } w_a^{n_a} w_b^{n_b} w_c^{n_c} w_d^{n_d} 
q^{\# paths}
\EN
where $w_a$($w_c$) and $w_b$($w_d$) are
the Boltzmann weights probabilities of right and left
mirrors(rotators) and $n_i(i=a,b,c,d)$ are the
the number of weights  in a given
configuration. Setting $q=1$ we recover the Lorentz gas model.

In the absence of rotators ($w_c=w_d=0$), the partition function (1) 
can be interpreted as a graphical representation of the $q^2$-state
Potts model, and therefore the algebraic structure underlying the mirror
collision rules is the standard TL algebra. This is no longer valid
if rotators are present, because a particle hitting a rotator does
not behave exactly the same when it hits a mirror. In fact, the
algebra that mimics the behaviour of rotators is as follows. We
associate to right and left rotators the elements $R_i$ and $L_i$, acting
on sites $i$ and $i+1$ of a quantum spin
chain of size $L$. We find that $R_i$ and $L_i$ are generators of
the following associative algebra
\EQ
R_i^2= qR_i,~~ L_i^2=qL_i,~~ [R_i,L_i]=0 
\EN
\EQ
L_i R_{i\pm1}L_i=L_i,~~ R_iL_{i\pm1}R_i=R_i,~~[R_i,R_{i\pm1}]=[L_i,L_{i\pm1}]=0
\EN
\EQ
[L_i,L_j]=[R_i,R_j]=[R_i,L_j]=0 
\hspace{0.3cm} |i-j| \geq 2
\EN

These relations can be seen as two commuting TL algebras whose generators
alternate between $R_i$ and $L_i$ depending on the parity of the sites. 
This algebra has to be read in conjunction with the TL operator $E_i=R_iL_i$
and the identity $I_i$, 
which in the Lorentz gas corresponds to left and right mirrors, respectively.
It is interesting to note that this algebra also allows us to
define the braid operator
$b_i= R_i +L_i -e^{\theta} E_i - e^{-\theta}I_i$ and its inverse
$b_i^{-1}= R_i +L_i -e^{-\theta} E_i - e^{\theta}I_i$, where
the  parameter $\theta$ is related to the fugacity by
$q=2 \cosh(\theta)$. The real surprise  is the fact that 
the braid-monoid operators $b_i$, $b_i^{-1}$ and 
$E_i$ satisfy 
the Birman-Wenzel-Murakami(BWM) algebra \cite{BWM}. More precisely, this 
connection occurs when the two independent parameters $\sqrt{Q}$ and
$c$ of the BWM algebra (we are using notations of ref.\cite{WA}) 
lie on the
curve $\sqrt{Q}=4 \cosh^2(\theta)$ and $c= -e^{3 \theta}$. 
However,
in the context of the Yang-Baxter equation, we find that
the ``rotator'' algebra (2-4) is richer than
the  BWM algebra(on the  above curve), since the former gives us
an extra integrable manifold.

To make further progress we look for possible integrable manifolds
for the Boltzmann weights $w_a$, $w_b$, $w_c$ and $w_d$. This not only
allows us to establish a theoretical framework to study the
Lorentz lattice gas but also prompts us to make exact predictions 
for its diffusive behaviour. The basic idea
is to search for solutions of the Yang-Baxter equation by Baxterizing
the Ansatz $w_a I_i +w_b E_i +w_c R_i +w_d L_i$ with the help of 
the algebraic relations (2-4). Here we omit technical details, presenting
only the final results. We find three different 
classes of integrable manifolds 
and they are given by
\EQ
(I)~~ w_a(\lambda)=\frac{\sinh(\theta-\lambda)}{\sinh(\lambda)},~ w_b(\lambda)=
1/w_a(\lambda),~ w_c(\lambda)=w_d(\lambda)=1
\EN
\EQ
(II)~~ w_a(\lambda)=\frac{\sinh(\theta-\lambda)}{\sinh(\lambda)},~
w_b(\lambda)=-\frac{\cosh(\theta-\lambda)}{\cosh(2\theta-\lambda)},~ 
w_c(\lambda)=w_d(\lambda)=1
\EN
\EQ
(III)~~ w_a(\lambda)=\frac{q}{e^{2 \lambda}-1},~
w_b(\lambda)=\frac{qe^{2 \lambda}}{q^2-1 -e^{2 \lambda}},~
w_c(\lambda)=1, w_d(\lambda)=0 
\EN
where $\lambda$ is the spectral parameter of the Yang-Baxter equation.
The third  solution  also admits $w_c(\lambda)=0$ and $w_d(\lambda)=1$,
since the algebra (2-4) is invariant under the exchange of left and
right generators.

In order to interpret these solutions it is more appealing to work with
a specific representation for the generators $R_i$ and $L_i$. In the
language of quantum spin chains, we find that these
generators can be written as
\EQ
R_i = \sigma_i^{+} \tau_{i+1}^{-} + \sigma_i^{-} \tau_{i+1}^{+} +
\frac{\cosh(\theta)}{2}( 1- \sigma_i^{z} \tau_{i+1}^{z}) +
\frac{\sinh(\theta)}{2}( \tau_{i+1}^{z}- \sigma_{i}^{z}) 
\EN
\EQ
L_i = \tau_i^{+} \sigma_{i+1}^{-} + \tau_i^{-} \sigma_{i+1}^{+} +
\frac{\cosh(\theta)}{2}( 1- \tau_i^{z} \sigma_{i+1}^{z}) +
\frac{\sinh(\theta)}{2}( \sigma_{i+1}^{z}- \tau_{i}^{z}) 
\EN
where $\{ \sigma_i^{\pm}, \sigma_i^{z} \}$ and $\{ \tau_i^{\pm}, \tau_i^{z} \}$
are two commuting sets of Pauli matrices acting on the sites of a lattice
of size $L$.

Considering this representation, it is not difficult to
see that solution I behaves similarly as 
two decoupled  $6$-vertex models  and therefore 
the underlying symmetry is based
on the $D_2$ Lie algebra \cite{JI}. 
The second solution corresponds to
a fine tuned coupled $6$-vertex models  through their (total) energy-energy
interaction and its 
quantum group symmetry, 
after a
canonical transformation, can be related 
to the twisted  $A_3^2$ Lie algebra \cite{JI}. The
third solution is clearly asymmetric in the spin variables and therefore
cannot be interpreted in terms of the BWM algebra. This solution
has also a number of unusual properties. For instance,
the asymptotic braid limits
$R(\pm \infty)$  are not invertible and 
for the special values $q =\pm 1$  we find a  connection to a 
$singular$ point  
($t=0$ in the notation
of \cite{BWM})
of the Hecke algebra. In the context of spin models, we have
verified that the Boltzmann weights (7) can be related to an
integrable non-self-dual manifold \cite{HP} of generalized Potts models
introduced by Domany and Riedel \cite{DR} to model the adsorption
of molecules on a crystal surface.

We  now turn our attention to the Lorentz lattice gas. For $q=1$ only
the first two solutions turns out to be physically meaningful, since
all the Boltzmann weights can be made positively definite. Thus,
we shall
concentrate our efforts to study the critical behaviour of these two
solutions with the expectation of making
predictions for the geometrical 
scaling behaviour of closed particle paths
in the Lorentz lattice gas. To this end, we study these models with
generalized boundary conditions, in order to assure that closed loops on the
cylinder pick up the correct Boltzmann weights. In general, these
are twisted boundary conditions defined by
\EQ
\sigma_{L+1}^{\pm}= e^{\pm i \psi_1} \sigma_1^{\pm},~
\tau_{L+1}^{\pm}= e^{\pm i \psi_2} \tau_1^{\pm},~
\sigma^{z}_{L+1}= \sigma^{z}_{1},~
\tau^{z}_{L+1}= \tau^{z}_{1}
\EN
where $\psi_1$ and $\psi_2$ are arbitrary angles.

The critical behaviour of solution I is that of two decoupled $XXZ$ 
spin chains and consequently the results for the central charge and
anomalous dimensions can be read off directly from previous work in
the literature \cite{BAT}.  This system is
critical in the regime $ q \in [2,-2]$ and  
it is more convenient to parametrize the fugacity 
by $q=2 \cos(\gamma)$. 
It turns out that the
effective central charge behaviour is 
\EQ
c= 2 -\frac{3}{x_p(\gamma)} \sum_{i=1}^{2} [\psi_i/2\pi]^2
\EN
while the conformal dimensions are given by
\EQ
X_{n_1,n_2}^{m_1,m_2} = \sum_{i=1}^{2} \left [ x_p(\gamma) n_i^2 
+\frac{1}{4 x_p(\gamma)}(m_i +\psi_i/2\pi)^2 \right ]
\EN
where $n_i$ and $m_i$ are integers representing the spin-wave and the
vortex excitations of two decoupled Coulomb gas having the
same 
radius amplitude $x_p(\gamma)=\frac{\pi-\gamma}{2 \pi}$. The
finite-size corrections (11,12) are measured relative to the 
ground state of two $periodic$ $XXZ$ models \cite{BAT}.

The second solution corresponds to two nontrivialy coupled 
$XXZ$ models and it is possible to show that its Bethe Ansatz 
solution is formally related to that of the $A_3^2$ vertex model \cite{RE}.
The  critical properties 
of the latter model  has only been partially studied
in the literature \cite{DEV} and in a region that exclude
the Lorentz lattice gas itself. We remark that here we
have to perform the 
calculations in the presence of the seam $\psi_1$ and $\psi_2$, 
since they plays 
a crucial role in the underlying critical behaviour. 
We find that the eigenvalues
of the Hamiltonian associated to the second solution are given by
\EQ
E_{II}(L)= \sum_{j=1}^{r_1} \frac{2 \sin^2(\gamma)}
{\cos(\gamma)
- \cosh(\lambda^{(1)}_j) }
\EN
and the corresponding Bethe Ansatz equations are
\EQ
\left[
\frac{\sinh(\lambda^{(1)}_{j}/2 -i\gamma/2)}
{\sinh(\lambda^{(1)}_{j}/2 +i\gamma/2)}
\right]^{L} =-e^{i\psi_1}
\prod_{k=1}^{r_1} 
\frac{\sinh(\lambda^{(1)}_{j}/2-\lambda^{(1)}_{k}/2 -i\gamma)}
{\sinh(\lambda^{(1)}_{j}/2-\lambda^{(1)}_{k}/2 +i\gamma)}
\prod_{k=1}^{r_2} 
\frac{\sinh(\lambda^{(1)}_{j}-\lambda^{(2)}_{k} +i\gamma)}
{\sinh(\lambda^{(1)}_{j}-\lambda^{(2)}_{k} -i\gamma)}
,j=1,\cdots,r_1
\EN
\EQ
e^{i(\psi_1-\psi_2)}\prod_{k=1}^{r_1} 
\frac{\sinh(\lambda^{(2)}_{j}-\lambda^{(2)}_{k} +i2\gamma)}
{\sinh(\lambda^{(2)}_{j}-\lambda^{(2)}_{k} -i2\gamma)} =
-\prod_{k=1}^{r_2} 
\frac{\sinh(\lambda^{(2)}_{j}-\lambda^{(1)}_{k} + i\gamma)}
{\sinh(\lambda^{(2)}_{j}-\lambda^{(1)}_{k} - i\gamma)},~j=1,\cdots,r_2
\EN

The existence of the Bethe Ansatz solution allows us to calculate,
besides the thermodynamic limit, the dominant finite-size corrections
for the eigenvalues of the Hamiltonian. For a conformally invariant system,
these finite-size effects can be directly related to the central charge
and scaling dimensions \cite{CA,CA1}, providing us a way to study
the universality class of solution II. An essential step here is
to determine the nature of the Bethe Anstaz roots 
governing the ground state properties.
In the regime we are interested, i.e. near $\gamma \sim \pi/3$, we find
a mixture between one-string and two-string type  of solutions. More
precisely, the complex roots structure are given by
\EQ
\lambda_j^{(1)} =\xi_j^{(1)} \pm i(\pi/2-\gamma) +O(e^{-L}),~ 
\lambda_j^{(2)} =\xi_j^{(2)} + i\pi/2 
\EN
where $\xi_j^{(a)}$ are real numbers.   

In order to determine the finite-size corrections for the ground state,
we have numerically solved the Bethe Ansatz equations for several values
of the lattice size up to $L \sim 40$. In table 1 we present our
estimates for the effective central charge for general boundary
conditions. Surprisingly, the behaviour is precisely the same of that
given in formula (11) which suggests that the coupling between
the two $XXZ$ models becomes asymptotically 
irrelevant 
near $q=1$. To give extra support to this scenario we have
also investigated the low-lying spin-wave excitations. In table 2 we show
some estimates for the exponents $X_{1,0}^{0,0}$ and $X_{1,1}^{0,0}$
and they are in accordance with the conformal dimensions (12).
More generally, we verified that the critical behaviour for role
region 
$q \in [0,\sqrt{2})$ is given in terms of two decoupled Coulomb-gas 
models. However, in the regime
$q \in [\sqrt{2},2]$ the coupling between the two $XXZ$ models becomes
relevant, and the criticality is governed by 
a $c=3$ conformal field theory. We note that in
the first branch all the
Boltzmann weights probabilities can
be positively defined while in the second one either $w_b$ or $w_a$ is
necessarily negative. This is perhaps the physical reason behind these
two different $antiferromagnetic$ critical behaviour.
Finally, we remark that our results 
complement and correct an early calculation in the literature \cite{DEV}. 

We now have the basic ingredients to study the scaling 
properties of closed trajectories in the Lorentz lattice gas model.
From our previous analysis we conclude that the critical behaviour,
for both manifolds I and II, is given in terms of two decoupled 
Coulomb-gas in the presence of the background charges $\psi_1=\psi_2=
2 \gamma =2 \pi/3$. Here we are interested in correlators that
measure the probabilities that two points on the lattice separated
by $r$ belong to the same loop \cite{NI}. For large distances we
expect the algebraic decay 
\EQ
< \Phi_l(r) \Phi_l(0) > \sim r^{-2 X_l}
\EN
where $X_l$ is the scaling dimension of the conformal operator
$\Phi_l(r)$ \cite{DU}. Equipped with equations (11) and (12) we can
now calculate the dimensions $X_l$ for the $q=1$ model. The probability
that $l=n_1+n_2$ loop segments meet in the neighbourhood of
two points in the lattice is associated to the conformal dimensions
of the spin-wave excitations $n_1$ and $n_2$ with null vortex charges 
$m_1=m_2=0$ \cite{NI,DU}. Due to the background 
charges, we also have to subtract the
variation of the true ground state for the Lorentz lattice gas. We conclude
that the value of these dimensions is
\bear
X_{n_1,n_2} & =& \sum_{i=1}^{2} \left [ 
x_p(\pi/3) n_i^2  -\frac{(1/3)^2}{4x_p(\pi/3)}(1-\delta_{n_i,0}) \right ]
\nonumber\\
& =& \sum_{i=1}^{2} \left [ 
\frac{n_i^2}{3}  -\frac{(1-\delta_{n_i,0})}{12} \right ]
\ear

The scaling behaviour of single paths is governed by the 
lowest conformal dimension $X_{1,0}=X_{0,1}=1/4$, predicting a  
fractal dimension \cite{Zif} $d_f =2 -1/4=7/4$. This is the same
value found 
in numerical 
simulations of a fully occupied lattice of either mirrors or
rotators \cite{Zif,CO2}. Therefore, our
result brings  an extra theoretical support to the fact that 
the scaling behaviour of single paths does not depend 
whether the scatterers
are mirrors or rotators. However, the situation changes
drastically when we consider the probability for
multi loops $l \geq 2$.
In fact, in the case of mixed
rotator-mirror model the second lowest
dimension occurs in the sector $n_1=n_2=1$, and the
corresponding fractal dimension is $d_f=2 -1/2=3/2$. This value
is double of that expected $d_f=2 -5/4=3/4$ when 
we have only mirrors \cite{DU}. In general, 
formula (18) leads us to conclude that
the multi loops scaling behaviour is
sensible to the details of the scattering mechanism.
Finally, we
remark that our findings are valid for rather distinct manifolds,
suggesting that $w_c=w_d$ is indeed a critical surface in the Lorentz
gas model.

In summary, we have introduced a rather rich variant of
the Temperley-Lieb algebra which generates three classes of
solutions of the Yang-Baxter equation. We have applied the first
two of them to study the diffusion behaviour of a Lorentz lattice
gas whose scatterers are 
either mirrors or rotators. The third manifold seems to
be relevant to the study of phase transition in adsorbed
films\cite{DR}, and we hope to investigate   
its critical properties 
in a future publication. 
Finally, we remark that after this paper has been written we noted
the very recent work \cite{PD} in which the $third$ manifold and its
generalizations were obtained via Baxterization of Fuss-Catalan algebras.

\section*{Acknowledgements}
This work was supported by  FOM (Fundamental Onderzoek der Materie) 
and Fapesp ( Funda\c c\~ao
de Amparo \`a Pesquisa do Estado de S. Paulo) and Cnpq (Conselho Nacional
de Desenvolvimento Cient\'ifico e Tecnol\'ogico).

\newpage
\underline{Table 1}: Finite size sequences for the 
extrapolation  of the effective central
charge for $\gamma=\pi/3$.

\begin{table}
\begin{center}
\begin{tabular}{|c|c|c|c|} \hline
     $L$  &$\psi_1=\psi_2=\pi/3$  &$\psi_1=\psi_2=\pi/4$ & $\psi_1=\pi/3$, $\psi_2=\pi/4$   \\ \hline\hline
8 & 1.55667 & 1.78890 & 1.67253  \\ \hline
16 & 1.51857 & 1.63017 & 1.63017  \\ \hline
24 & 1.51021 & 1.73159 & 1.62083  \\ \hline
32 & 1.50683 & 1.72737 & 1.61705  \\ \hline
40 & 1.50506 & 1.72516 & 1.61507 \\ \hline
Extr. & 1.5001(2)  & 1.7186(2) & 1.6092(2) 
\\ \hline
\end{tabular}
\end{center}
\end{table}

\underline{Table 2}: Finite size sequences for the extrapolation 
of the  spin-wave  anomalous dimensions $X_{1,0}^{0,0}$ and
$X_{1,1}^{0,0}$ for periodic boundary conditions.

\begin{table}
\begin{center}
\begin{tabular}{|c|c|c|c|c|} \hline
     $L$  &$X_{1,0}^{0,0}(\gamma=\pi/3)$  &$X_{1,1}^{0,0}(\gamma=\pi/3)$ & 
$X_{1,0}^{0,0}(\gamma=\pi/3.5)$ & $X_{1,1}^{0,0}(\gamma=\pi/2.5)$   \\ \hline\hline
8 &0.344610  &0.640435  & 0.404331  & 0.599395\\ \hline
16 &0.337923  &0.653822  & 0.392503 & 0.599587 \\ \hline
24 &0.336174  &0.658341  & 0.387451 & 0.599756 \\ \hline
32 &0.335383  &0.660548  & 0.384344 & 0.599839 \\ \hline
40 &0.334934  &0.661844  & 0.382155 & 0.599890\\ \hline
Extr. & 0.33330(2)  & 0.66665(2) &0.3575(3) & 0.6000(2) 
\\ \hline
\end{tabular}
\end{center}
\end{table}

\end{document}